\documentclass[sigplan,screen,nonacm]{acmart}

\setcopyright{none}
\usepackage{amsmath}
\usepackage{algorithmic}
\usepackage{graphicx}
\usepackage{textcomp}
\usepackage{xcolor}
\usepackage{listings}
\usepackage{xspace}
\usepackage{multirow}
\usepackage{subcaption}
\usepackage{enumitem}
\usepackage{microtype}
\usepackage{balance}
\usepackage{float}
\newcommand{\sqlk}[1]
{\text{\ttfamily\footnotesize\bfseries#1}}

\lstdefinelanguage{datalog}{
  morekeywords={input,output,decl,number,symbol,unsigned,float,union,NULL,local,multi},,
  morecomment=[l]{//},
  sensitive=true,
  morestring=[b]"
}
\begin{document}

\title{Compiling Linear Datalog to SQL for Program Analysis}

\author{Amir Shaikhha}
\email{amir.shaikhha@tu-darmstadt.de}
\affiliation{
  \institution{TU Darmstadt, Hessian.ai, University of Edinburgh}
  \country{Germany, United Kingdom}
}

\author{Anna Herlihy}
\email{anna.herlihy@epfl.ch}
\affiliation{
  \institution{EPFL}
  \country{Switzerland}
}

\author{Hung Ngo}
\email{hung.ngo@relational.ai}
\affiliation{
  \institution{RelationalAI}
  \country{United States}
}

\begin{abstract}
Datalog is a declarative query language that has proven highly effective for expressing static program analyses. Although Datalog has deep roots in database theory, most recent advances have largely emerged from the programming languages and compiler communities, with systems such as Soufflé. In contrast, modern relational engines have made significant progress in optimizing recursive SQL.

This paper revisits the connection between Datalog and relational databases,  advocating recursive SQL as a backend for Datalog evaluation. We present a compilation framework that translates Datalog programs, particularly those in the Linear Datalog fragment, into equivalent recursive SQL queries. To bridge the gap between Datalog and SQL, the compiler routes every program through an intermediate language called Midlog. The compiler additionally recovers functional dependencies from the program and exposes them as schema keys, unlocking the engine’s standard query optimizations.
This approach enables existing database engines to execute a broad class of program analyses, outperforming the Soufflé engine by up to an order of magnitude on the Umbra backend. Umbra achieves a geometric-mean speedup of 5.46$\times$ at 8 threads, whereas DuckDB is competitive with Soufflé single-threaded and is slower at 8 threads (geometric-mean speedup of 0.68$\times$). Furthermore, the generated SQL is portable; it runs on seven database systems without any engine modification. Our results highlight what the relational engines require to fully support Datalog for large-scale program analysis.
\end{abstract}
\maketitle

\section{Introduction}
Datalog is a declarative programming language popular in program analysis: many static analyses, such as points-to analysis, call-graph construction, and dataflow analysis, are fixed-point computations.
A direct implementation of such analyses in a general-purpose language requires encoding how the fixed point is reached (worklists, dependency tracking, iteration order). In Datalog, it is a set of recursive rules describing how to derive new facts from known facts, with a bottom-up engine computing the least fixed point.
This combination of expressiveness and efficiency has made Datalog an attractive choice for program analysis tasks in both academia and industry ~\cite{souffle,madsen2016flix,aref2025rel,aref2015design}.

Despite its database-theory roots, most academic development of Datalog has happened outside the database community. State-of-the-art systems such as Soufflé~\cite{souffle} are standalone compilers translating Datalog into optimized C++, bypassing relational engines entirely, even though Datalog is relational algebra extended with a fixed-point operator~\cite{abiteboul1995foundations}.

Relational engines (RDBMS) have meanwhile made significant progress in recursion support. The SQL:1999 standard introduced \emph{recursive common table expressions} (CTEs), which expose the entire recursive computation to the query optimizer, enabling global rewrites, indexing, cost-based optimization, and parallel execution~\cite{duckdb,sql1999}.

These two lines of work have evolved in isolation. Datalog compilers extract performance through specialization, but cannot reuse the indexing, optimization, and parallelism of RDBMS. RDBMS support recursive queries, but lack Datalog's concise abstractions and expressiveness. We target both axes, compiling Datalog into recursive SQL so that RDBMS serve as efficient Datalog backends without custom runtimes.

In this paper, we propose DLSQL\xspace, a compiler that translates \emph{Linear Datalog}, a well-defined fragment of Datalog in which each recursive rule contains at most one recursive predicate in its body, into recursive SQL. Thanks to this restriction, recursion maps directly to SQL's recursive CTEs.

Even within Linear Datalog\xspace, a substantial mismatch with recursive SQL remains. Midlog\xspace is an intermediate language that bridges this gap and is designed around two fundamental differences. First, despite the support for mutual recursion in SQL:1999, no engine we tested implements it (except for MariaDB); thus, unlike Datalog, SQL in practice does not support mutual recursion~\cite{herlihy2026lir}. Second, SQL employs a fundamentally different scoping discipline than Datalog. 

We make the following contributions:
\begin{itemize}[topsep=2pt,itemsep=1pt,leftmargin=*]
  \item We identify \emph{Linear Datalog} as an expressive subset of Datalog and express reduced kernels of five canonical static analyses, spanning call-graph construction, borrow checking, and points-to/escape/dataflow analysis  (Section~\ref{sec:linear-datalog}).
  \item We propose Midlog\xspace, an intermediate language that resolves the two mismatches between Linear Datalog\xspace\ and recursive SQL:  mutual recursion and scoping (Section~\ref{sec:midlog}).
  \item We present the DLSQL\xspace\ compiler pipeline, which routes every program through Midlog\xspace\ in four phases (Section~\ref{sec:compiler}).
  \item We develop two optimizations: a multi-head elimination removing unnecessary mutual recursion, and a functional-dependency pass recovering schema keys from the input data, enabling the engine to optimize further (Section~\ref{sec:opt}).
  \item We evaluate DLSQL\xspace\ on five static program-analysis workloads across seven database engines and against four state-of-the-art Datalog systems: Soufflé, FlowLog, Logica, and RecStep. On Umbra, DLSQL\xspace\ achieves a geometric-mean speedup of 5.46$\times$ over Soufflé at 8 threads, and up to 14.3$\times$ on individual benchmarks. On DuckDB, it reaches 0.96$\times$ single-threaded and 0.68$\times$ at 8 threads. Furthermore, we study portability across engines, the impact of both optimizations, and scaling with input size (Section~\ref{sec:exp}).
\end{itemize}

\section{Background}\label{sec:background}

This section covers the two languages that meet in this paper: Datalog as used by program-analysis frameworks, and recursive SQL as supported by modern RDBMS. Throughout, we treat the database engine as a black-box recursive-query evaluator: our compiler emits recursive \textsc{CTE}\xspace{}s in the SQL:1999 standard and relies on the engine's own evaluation strategy.

\subsection{Datalog}

A Datalog program is a finite set of \emph{rules} of the form
\[
H(\vec{t}) \;\mbox{:-}\; L_1, \ldots, L_n.
\]
where $H(\vec{t})$ is the \emph{head} atom and the $L_i$ are \emph{body literals}.
An \emph{atom} applies a \emph{relation} (or \emph{predicate}) to one or more \emph{terms}, each a variable or a constant.
A literal is a positive atom $R(\vec{t})$, a negated atom $\neg R(\vec{t})$, or a comparison $t \mathbin{\theta} t'$ with $\theta \in \{=,\neq,<,\le,>,\ge\}$. Read declaratively, the rule states that $H$ holds for every assignment of variables under which all body literals hold. The comma denotes conjunction.

A relation is either \textsc{EDB}\xspace (extensional) holding stored input facts, or \textsc{IDB}\xspace (intensional) derived by the rules. A program is evaluated bottom-up: starting from the \textsc{EDB}\xspace facts, the rules are applied repeatedly until no new fact can be added.

Because predicates may depend on one another recursively, evaluation is ordered by the \emph{predicate dependency graph}, which has a node per predicate and an edge $R \to H$ whenever $R$ occurs in the body of a rule defining $H$. A \emph{strongly connected component} (\textsc{SCC}\xspace) of this graph is a maximal set of predicates reachable from one another.

Negation is permitted only in stratified form: a negated atom never refers to a predicate in its own \textsc{SCC}\xspace, so it lies in an earlier, fully evaluated \textsc{SCC}\xspace. 
The model of the entire stratified program is constructed by computing the least fixed point of each \textsc{SCC}\xspace individually, in a topological order of the \textsc{SCC}\xspace dependency graph, with the earlier \textsc{SCC}\xspace{}s held fixed. The resulting model is the program's perfect model~\cite{przymusinski1988declarative,abiteboul1995foundations}.

Finally, a rule is \emph{safe} when every variable in its head, in a negated atom, or in a comparison also appears in some positive body atom, guaranteeing that every derived fact is bound to constants and that the result is finite~\cite{ceri1989you,abiteboul1995foundations}.

Most Datalog systems extend this core with schema and I/O declarations. We adopt Soufflé's surface syntax~\cite{souffle}: a relation is declared with \texttt{.decl} (named, typed columns), tied to a CSV file with \texttt{.input}, and exposed as a query result with \texttt{.output}. Figure~\ref{fig:datalog-tc} shows transitive closure in this syntax.

\begin{figure}[t]
\begin{minipage}[t]{.35\columnwidth}
\vspace{-0.2cm}
\begin{lstlisting}[language=datalog]
.decl edge(x:number, 
            y:number)
.decl tc(x:number, 
          y:number)
.input edge
.output tc
tc(x,y) :- edge(x,y).
tc(x,z) :- tc(x,y), 
             edge(y,z).
\end{lstlisting}
\vspace{-0.6cm}
\subcaption{Soufflé's Datalog.}
\label{fig:datalog-tc}
\end{minipage}
\begin{minipage}[t]{.64\columnwidth}
\begin{quote}
\sqlk{WITH RECURSIVE} \textit{tc}(x, y) \sqlk{AS} (\\
\hspace*{1em}\sqlk{SELECT} x, y \sqlk{FROM} \textit{edge}\\
\hspace*{1em}\sqlk{UNION}\\
\hspace*{1em}\sqlk{SELECT} \textit{tc}.x, \textit{edge}.y \\
\hspace*{1em}\sqlk{FROM} \textit{tc}, \textit{edge}\\
\hspace*{1em}\sqlk{WHERE} \textit{tc}.y = \textit{edge}.x\\
)\\
\sqlk{SELECT} * \sqlk{FROM} \textit{tc}
\end{quote}
\subcaption{Recursive SQL.}
\label{fig:sql-tc}
\end{minipage}
\vspace{-0.3cm}
\caption{Transitive closure in Datalog and SQL.}
\label{fig:query-tc}
\vspace{-0.4cm}
\end{figure}

\subsection{Recursive Common Table Expressions}

Recursive queries entered the SQL standard in SQL:1999 as \emph{recursive common table expressions} (\textsc{CTE}\xspace{}s)~\cite{sql1999,eisenberg1999sql}, written
\begin{quote}
\sqlk{WITH RECURSIVE} $R(\vec{c})$ \sqlk{AS} ($Q_{\text{base}}$ \sqlk{UNION} $Q_{\text{step}}$) \sqlk{SELECT \ldots},
\end{quote}
where $Q_{\text{base}}$, the ``base-case'', does not reference $R$, and $Q_{\text{step}}$, the ``recursive-case'', references $R$. $Q_{\text{step}}$ can itself contain \sqlk{UNION} branches, and depending on the engine, nested \sqlk{WITH} expressions. Iteration proceeds until a fixed point is reached.
The standard imposes \emph{linear recursion} and restricts where the recursive reference may occur: $R$ must appear at most once per \sqlk{UNION} branch of $Q_{\text{step}}$ (including within its \sqlk{WITH} blocks). \sqlk{UNION} provides set semantics over the iterations. The transitive-closure Datalog program shown earlier corresponds to the SQL query shown in Figure~\ref{fig:sql-tc}.

Note that SQL:1999 does define mutually recursive definitions~\cite{sql1999}. However, of the SQL engines we evaluate, only MariaDB accepts them, so this constraint is implementation support rather than the standard itself.

Behind this surface syntax, modern engines apply standard bottom-up semi-naive execution: the base case seeds a working table, the recursive case re-runs against it, and the difference against the known tuples feeds the next round~\cite{bancilhon1985naive,balbin1987generalization}. From the query writer's perspective this is invisible: a \sqlk{WITH RECURSIVE} block expresses the desired least-fixed-point relation, and the engine evaluates it efficiently.

This separation of concerns is what our compiler exploits. By emitting only plain \sqlk{WITH RECURSIVE} blocks, DLSQL\xspace\ inherits whatever evaluation and optimization strategies the host engine provides. The same compiler output runs on DuckDB~\cite{duckdb}, Umbra~\cite{neumann2020umbra}, and Hyper~\cite{kemper2011hyper}, and, after an inlining pass, on MariaDB, MySQL, and SQLite (cf. Section~\ref{sec:exp:portability}).

\section{Linear Datalog}\label{sec:linear-datalog}

\subsection{Syntax}\label{sec:syntax}

We define Linear Datalog\xspace as the fragment of Datalog in which every rule is \emph{linear}: at most one body literal refers to a predicate in the same \textsc{SCC}\xspace as the head. Every recursive rule therefore has the shape $H(\vec{t}) \;\mbox{:-}\; R(\vec{s}),\; C_1, \ldots, C_k$, where the recursive literal $R(\vec{s})$ is positive and in the same \textsc{SCC}\xspace as $H$, and each $C_i$ is an \textsc{EDB}\xspace, an \textsc{IDB}\xspace from an earlier \textsc{SCC}\xspace, or a comparison. Mutual recursion is allowed, since $R$ needs not to be equal to $H$, but only belong to the same \textsc{SCC}\xspace. 

\noindent\textbf{Why linearity matters.} Linearity is the recursion class that SQL:1999 admits: the recursive term may reference a recursive relation at most once~\cite{eisenberg1999sql}. Most relational engines implement a further-restricted form of it, i.e., no support for mutual recursion~\cite{herlihy2026lir}. Within the fragment, the entire fixpoint of an analysis compiles to a single recursive query that runs on unmodified engines (cf. Section~\ref{sec:exp:portability}).
Restricting to Linear Datalog\xspace thus captures the reduced analysis kernels of Section~\ref{sec:analyses}, while remaining portable; we return to what falls outside the fragment at the end of this section.

Prior theory establishes the expressibility correspondence between linear Datalog and SQL:1999 recursion. Our contribution is a compiler that makes it run by accounting for two practical challenges. First, although the SQL:1999 standard allows mutual recursion, most RDBMS do not support it. Second, Datalog's scoping discipline differs from SQL. Midlog\xspace's two rewrites (Sections~\ref{sec:midlog-normalization} and~\ref{sec:opt-multihead}) close exactly these gaps.

\subsection{Program Analyses Expressible in Linear Datalog}\label{sec:analyses}

A variety of canonical static analyses can be expressed in Linear Datalog\xspace, sharing a common shape: the analysis state is a transitive-closure-like relation, propagated step by step through non-recursive context relations. We illustrate with five analyses, all used as benchmarks in Section~\ref{sec:exp}, starting with points-to analysis.
These five benchmarks are the recursive cores of real analyses rather than full-fledged analyses. We state below what each kernel drops relative to its source, and we cross-validate the output of every kernel against Soufflé on every engine, where it completes (cf. Section~\ref{sec:exp:setup}).

\begin{figure}[t]
\begin{lstlisting}[language=datalog,basicstyle=\ttfamily\scriptsize,aboveskip=2pt,belowskip=2pt]
// R1: an allocation in a reachable method
VarPointsTo(heap, var) :-
    AssignHeapAlloc(heap, var, method), Reachable(method).
// R2: assignment copies the points-to set
VarPointsTo(heap, to) :-
    Assign(from, to), VarPointsTo(heap, from).
// R3: store to a static field
StaticFieldPointsTo(heap, fld) :-
    VarPointsTo(heap, from), Reachable(method),
    StoreStaticField(from, fld, method).
// R4: load from a static field
VarPointsTo(heap, to) :-
    Reachable(method), LoadStaticField(fld, to, method),
    StaticFieldPointsTo(heap, fld).
\end{lstlisting}
\vspace{-0.4cm}
\caption{Four of \texttt{varpointsto}'s ten rules; each has at most one positive recursive reference, so the program is linear.}
\label{fig:pts-rules}
\vspace{-0.3cm}
\end{figure}

\begin{figure}[t]
\vspace{-0.23cm}
\centering
\footnotesize
\begin{tabular}{@{}r@{\;}l@{\hspace{1.2em}}l@{}}
\multicolumn{2}{c}{\textbf{Java statement (in \texttt{main})}} & \textbf{Extracted \textsc{EDB}\xspace fact} \\
\hline
(1) & \texttt{Object x = new Object();} & \textit{AssignHeapAlloc}($h_1$, \texttt{x}, \texttt{main}) \\
(2) & \texttt{Object w = new Object();} & \textit{AssignHeapAlloc}($h_2$, \texttt{w}, \texttt{main}) \\
(3) & \texttt{Object y = x;}            & \textit{Assign}(\texttt{x}, \texttt{y}) \\
(4) & \texttt{C.s = y;}                 & \textit{StoreStaticField}(\texttt{y}, \texttt{s}, \texttt{main}) \\
(5) & \texttt{Object z = C.s;}          & \textit{LoadStaticField}(\texttt{s}, \texttt{z}, \texttt{main}) \\
\end{tabular}
\vspace{-0.3cm}
\caption{A Java fragment and the \textsc{EDB}\xspace tuples that the front end extracts from each statement, one tuple per statement.}
\label{fig:pts-java}
\vspace{-0.4cm}
\end{figure}

\noindent\textbf{Points-to analysis.} Points-to analysis determines, for each variable, the set of heap objects it may refer to at runtime; the flagship Datalog benchmark is the \textit{VarPointsTo} relation from Doop~\cite{bravenboer2009strictly}.
Here we use a reduced variant of Doop's micro workload, which is already context-insensitive. In addition, our variant no longer computes reachable methods or the points-to information of instance fields and array indices inside the fixpoint. In the original Doop micro, their corresponding rules are non-linear because they match two \textsc{IDB}\xspace{}s in one body. Changing these to consume precomputed input relations makes their rules linear.

Each rule propagates a heap object along an assignment, cast, field load, array load, or virtual-call resolution. The full \textit{VarPointsTo} relation is defined by ten rules and is linear. Figure~\ref{fig:pts-rules} shows four of the ten rules, taken directly from this Doop variant. The complete program appears in Figure~\ref{fig:varpointsto-full}.

To illustrate, Figure~\ref{fig:pts-java} shows a Java fragment in which the object allocated on line 1 flows through the static field \texttt{C.s}, forcing the analysis through the heap, together with the single \textsc{EDB}\xspace tuple extracted per statement; heap objects are labeled by allocation site ($h_1, h_2$), and the method \texttt{main} is assumed reachable (\textit{Reachable}(\texttt{main})).

\noindent\textbf{Call-graph construction.}
Call-graph construction determines which methods each call site may invoke.
Figure~\ref{fig:cha} shows the rules, which resolve a virtual call by class-hierarchy analysis (CHA) to every implementation of the called method in the receiver's declared type or a subtype of it (the non-recursive \textit{Subtype} and \textit{MethodImpl} relations), while statically-bound calls carry their target directly. \textit{ReachableMethod} and \textit{CallEdge} are mutually recursive: a method is reachable when a reachable method calls it, and a call edge is created only at a reachable call site. Every rule contains at most one \textsc{IDB}\xspace{} in the \textsc{SCC}\xspace $\{\textit{ReachableMethod}, \textit{CallEdge}\}$, so each rule is linear.

\begin{figure}[t]
\begin{lstlisting}[language=datalog,basicstyle=\ttfamily\scriptsize,aboveskip=2pt,belowskip=2pt]
// Entry methods are reachable.
ReachableMethod(m) :- Entry(m).
// Reachability propagates through discovered call edges.
ReachableMethod(ce) :- CallEdge(_, ce).
// Virtual calls: every impl in a subtype of the receiver.
CallEdge(cs, ce) :- ReachableMethod(ca),
    VirtualCall(cs, ca, declType, name, desc),
    Subtype(concrete, declType),
    MethodImpl(concrete, name, desc, ce).
// Statically-bound calls carry their callee directly.
CallEdge(cs, ce) :- ReachableMethod(ca), DirectCall(cs, ca, ce).
\end{lstlisting}
\vspace{-0.4cm}
\caption{Class-hierarchy call-graph construction (\texttt{cgcha}).}
\label{fig:cha}
\vspace{-0.4cm}
\end{figure}

\noindent\textbf{Escape analysis.} An object \emph{escapes} its allocating method if it is reachable from a static field, returned, passed as an actual argument, or used as a virtual-call receiver. Allocations that do not escape can then be stack-allocated rather than heap-allocated.
Figure~\ref{fig:escape} in the appendix gives the core rules of the Soufflé \texttt{escape} benchmark: \textit{HeapReachable} is a linear transitive closure over the heap graph, \textit{GlobalEscape} closes under it, \textit{MethodEscape} collects the ways an allocation escapes (including reachability from another escaping allocation), and \textit{CapturedAllocation} applies stratified negation against \textit{MethodEscape}. Each recursive rule references its own \textsc{SCC}\xspace's relation at most once, so the program is linear.

\noindent\textbf{Dataflow analysis.} A dataflow analysis propagates facts along a precomputed value-flow graph. Our \texttt{csda} benchmark performs null-value propagation over the context-sensitive value-flow graph of Graspan~\cite{graspan}, whose context-sensitivity is baked into the node IDs by cloning function bodies per calling context, reducing the analysis to plain reachability. \textit{NullNode} starts from the directly-null edges and propagates along the non-recursive \textit{Edge} relation, referencing itself once, so the recursion is linear:

\begin{lstlisting}[language=datalog]
NullNode(x, y) :- NullEdge(x, y).
NullNode(x, y) :- NullNode(x, w), Edge(w, y).
\end{lstlisting}

\noindent\textbf{Borrow checking.}
Borrow checking determines whether
a program ever uses a value after it has been moved. Polonius~\cite{polonius} reformulates Rust's borrow checker as a Datalog program. Figure~\ref{fig:polonius} in the appendix shows the recursive core of the \texttt{borrow} benchmark: \textit{ancestor\_path} is a linear transitive closure of the \textsc{EDB}\xspace \textit{child\_path} relation; \textit{path\_moved\_at} and \textit{path\_assigned\_at} propagate base \textsc{EDB}\xspace facts along \textit{ancestor\_path}; \textit{path\_maybe\_uninitialized\_on\_exit} flows move information along \textit{cfg\_edge} with stratified negation against \textit{path\_assigned\_at}; and a final non-recursive join reports each \textit{move\_error}. Every recursive rule contains exactly one positive recursive literal, so the program is linear.

\noindent\textbf{What falls outside the fragment.} Not every analysis fits Linear Datalog\xspace. A field-sensitive points-to analysis is representative: its instance-field store and load rules match two points-to facts in a single body, and Andersen-style inclusion-based points-to inherits the same non-linear closure rules. Three routes exist beyond the fragment: engine-specific extensions of recursive SQL, at the price of portability; an external driver iterating non-recursive queries to a fixpoint, as RecStep~\cite{recstep2019} and Logica~\cite{logica} do, at the price of hiding the fixpoint from the engine's optimizer; and compile-time rewrites into the fragment, via non-linear-to-linear transformations~\cite{sagiv1988optimizing} that are applicable to a limited class of non-linear programs. The first two routes forfeit the single-portable-query property that motivates our design.

\section{The Midlog\xspace Intermediate Representation}\label{sec:midlog}

Midlog\xspace is a rule-based intermediate language between Linear Datalog\xspace and recursive SQL. Like Datalog, a Midlog\xspace program is a set of rules, each defining one relation; unlike Datalog, a rule's body is a single \emph{union of conjunctions}, and a rule may nest other rules as lexically-scoped \emph{locals}~\cite{shaikhha2025hojabr}.
Midlog\xspace{} absorbs two structural mismatches between Datalog and SQL.

The first is mutual recursion: a Datalog program often defines several relations in terms of one another, forming a \emph{recursive group} or \textsc{SCC}\xspace of the predicate dependency graph. SQL:1999 defines mutually recursive query names within a \sqlk{WITH RECURSIVE} block; however, of the engines we evaluate only MariaDB implements them (cf. Section~\ref{sec:exp:portability}), so a group cannot be portably emitted as several recursive \textsc{CTE}\xspace{}s side by side.
Midlog\xspace resolves this by making one relation the recursive \emph{primary}; the remaining relations of the \textsc{SCC}\xspace become \emph{companions}, translated into non-recursive \textsc{CTE}\xspace{}s.

The second is scoping: SQL's \sqlk{WITH} blocks are lexically scoped and may nest, whereas Datalog uses a flat global namespace; Midlog\xspace nests the companions as locals inside the primary's rule, visible only to it.

\subsection{Syntax}\label{sec:midlog-syntax}

A Midlog\xspace program is a schema (inherited verbatim from the source program) plus a list of rules; Figure~\ref{fig:midlog-grammar} shows the grammar. Atoms are as in Datalog: positive accesses $R(\vec{t})$, negated accesses $\neg R(\vec{t})$, and comparisons $t \mathbin{\theta} t'$. Two properties distinguish Midlog\xspace from surface Datalog:

\begin{figure}[t]
\centering
\footnotesize
\begin{tabular}{r@{\;}c@{\;}l}
$\mathit{Rule}$     & ::= & $\mathit{Def} \;\;|\;\; \mathit{MultiDef}$ \\
$\mathit{Def}$      & ::= & $\mathit{Head}\;\sqlk{:-}\; \mathit{Local}^*\;\; \mathit{Body}$ \\
$\mathit{MultiDef}$ & ::= & $\sqlk{multi:}\;\; \mathit{Local}^*\;\; \mathit{Branch}^+$ \\
$\mathit{Branch}$   & ::= & $\mathit{Head}\;\sqlk{:-}\; \mathit{Body}$ \\
$\mathit{Local}$    & ::= & $\sqlk{local}\;\; \mathit{Def}$ \\
$\mathit{Head}$     & ::= & $\mathit{Name}(\mathit{Col}^*)$ \\
$\mathit{Body}$     & ::= & $\mathit{Conj}\;\; (\sqlk{union}\;\; \mathit{Conj})^*$ \\
$\mathit{Conj}$     & ::= & $\mathit{Atom}\;\; (\texttt{,}\;\; \mathit{Atom})^*$ \\
\end{tabular}
\vspace{-0.3cm}
\caption{Midlog\xspace grammar. Atoms are identical to Datalog.}
\label{fig:midlog-grammar}
\vspace{-0.3cm}
\end{figure}

\noindent\textbf{Pre-unioned bodies.} Several Datalog rules with the same head merge into a \emph{single} rule whose body is a \sqlk{union} of conjunctions, one per original rule, mirroring the output SQL: one \sqlk{WITH} entry per rule, one \sqlk{UNION} branch per conjunction.

\noindent\textbf{Lexically-scoped locals.} Mutually-recursive predicates become a single recursive primary with the companions nested as locals, compiled into non-recursive inner \sqlk{WITH} statements that wrap the primary's recursive case.
The $\mathit{MultiDef}$ form (\sqlk{multi:}) packages the heads of an \textsc{SCC}\xspace as they leave the Datalog-to-Midlog\xspace phase; Midlog\xspace normalization (Section~\ref{sec:midlog-normalization}) then rewrites every $\mathit{MultiDef}$ into a primary $\mathit{Def}$ with locals.

\subsection{Midlog\xspace Invariants}\label{sec:midlog-invariants}

Every Midlog\xspace program satisfies structural invariants on which the SQL emitter and the unparser rely; each holds by construction, following from the grammar of Figure~\ref{fig:midlog-grammar} or asserted as the compiler builds each program.

\begin{itemize}[leftmargin=*]
  \item[(I1)] \textbf{Unique heads per scope.} Head names are unique within a $\mathit{MultiDef}$'s branch list and the top-level rule list, so each head names exactly one definition, rendered as one uniquely-named \textsc{CTE}\xspace.
  \item[(I2)] \textbf{Bodies unioned per head.} All Datalog rules sharing a head merge into a single $\mathit{Body}$, one $\mathit{Conj}$ per original rule, so every relation compiles to exactly one \textsc{CTE}\xspace.
  \item[(I3)] \textbf{Canonical projection.} Each conjunction is $\alpha$-renamed so the enclosing head's declared column names appear directly as body variables; the emitter reads each \sqlk{SELECT}'s projection from the head alone.
  \item[(I4)] \textbf{Schema-derived columns.} A head's columns come from the Datalog \texttt{.decl} declaration when present, else from the first defining rule's head variables, so every emitted \textsc{CTE}\xspace has an explicit, stable column list (\sqlk{WITH}~$R(c_1,\dots)$).
\end{itemize}

\subsection{Normalized Midlog\xspace}\label{sec:normalized-midlog}
Normalized Midlog\xspace retains the grammar of Figure~\ref{fig:midlog-grammar}, except every top-level rule is a single-head $\mathit{Def}$, each multi-member \textsc{SCC}\xspace is a primary $\mathit{Def}$ with its companions as locals, and every $\mathit{Def}$ is at most self-recursive.
The invariants above continue to hold, and normalization additionally establishes:

\begin{itemize}[leftmargin=*]
  \item[(I5)] \textbf{Lexically-scoped locals.} A rule's locals are visible inside its body and to each other, and may reference the enclosing head; a companion can thus sit inside the primary's recursive case as an inner \sqlk{WITH}.
  \item[(I6)] \textbf{\textsc{SCC}\xspace hoisting.} The chosen primary $P$ is a top-level $\mathit{Def}$ owning the group's projections as locals; each projection referenced from outside the group, or output by the program, is additionally hoisted as a top-level sibling with the same head and body, reachable where the local copy is out of scope.
\end{itemize}

\noindent Figure~\ref{fig:midlog-cha} shows the initial and normalized Midlog\xspace representations of the CHA call-graph program of Figure~\ref{fig:cha}. The next section describes the producing pipeline, and Section~\ref{sec:opt-multihead} an optimization that, for \texttt{cgcha}, lifts a natural primary and avoids the synthetic combined relation (Figure~\ref{fig:midlog-cha-opt}).

\begin{figure*}[t]
\begin{minipage}{.38\textwidth}
\begin{lstlisting}[language=datalog,basicstyle=\ttfamily\footnotesize]
multi:
  ReachableMethod(m) :-
    Entry(m) union CallEdge(a1, m).

  CallEdge(cs, ce) :-
    ReachableMethod(ca),
    VirtualCall(cs, ca, declType, name, desc),
    Subtype(concrete, declType),
    MethodImpl(concrete, name, desc, ce)
      union
    ReachableMethod(ca), DirectCall(cs, ca, ce).
\end{lstlisting}
\subcaption{The initial Midlog\xspace form.}
\label{fig:midlog-cha-initial}
\end{minipage}
\begin{minipage}{0.61\textwidth}
\begin{lstlisting}[language=datalog,basicstyle=\ttfamily\footnotesize]
ReachableMethod_CallEdge(tag, c1, c2) :-
  local ReachableMethod(m) :- ReachableMethod_CallEdge(RM, m, c2).
  local CallEdge(cs, ce) :- ReachableMethod_CallEdge(CE, cs, ce).
  Entry(c1), c2 = NULL, tag = RM  union
  CallEdge(a1, c1), c2 = NULL, tag = RM union
  ReachableMethod(ca), VirtualCall(c1, ca, declType, name, desc),
  Subtype(concrete, declType),  MethodImpl(concrete, name, desc, c2), tag = CE
      union
  ReachableMethod(ca), DirectCall(c1, ca, c2), tag = CE.
ReachableMethod(m)         :- ReachableMethod_CallEdge(RM, m, c2).
CallEdge(cs, ce)       :- ReachableMethod_CallEdge(CE, cs, ce).
\end{lstlisting}
\subcaption{Normalized Midlog\xspace form. The $\mathit{tag}$ column distinguishes the two original branches.}
\label{fig:midlog-cha-normalized}
\end{minipage}
\vspace{-0.4cm}
\caption{Different Midlog\xspace representations of the CHA call-graph program (Figure~\ref{fig:cha}).}
\label{fig:midlog-cha}
\vspace{-0.3cm}
\end{figure*}

\noindent\textbf{Semantics.} Midlog\xspace inherits the semantics of stratified Datalog (the fixed-point semantics of Section~\ref{sec:background}), and a Midlog\xspace program denotes the same model as the Datalog program it normalizes, with locals considered as ordinary (scoped) \textsc{IDB}\xspace{} definitions.
The two rewrites that establish the normal form are equivalence-preserving under stated conditions: for the tagged rewrite (Section~\ref{sec:midlog-normalization}), the least fixpoint of the combined relation is the tag-disjoint union of the \textsc{SCC}\xspace's relations, with one tagged tuple per original tuple; the multi-head rewrite is sound under the conditions specified in Section~\ref{sec:opt-multihead}. A mechanized proof of both equivalences is future work.

\section{Compiling Linear Datalog\xspace to SQL}\label{sec:compiler}

DLSQL\xspace\ lowers a Linear Datalog\xspace program to a single recursive SQL query in four phases: (i) normalization of the Datalog source, (ii) translation to Midlog\xspace, (iii) Midlog\xspace{} normalization, and (iv) translation to a typed SQL IR, which a final unparser renders as an engine-portable query.
The compiler implements no fixed-point strategy of its own: it emits one \sqlk{WITH RECURSIVE} \textsc{CTE}\xspace per recursive group and lets the database engine perform iterative evaluation.

Throughout this section, we use the \texttt{cgcha} program of Figure~\ref{fig:cha} as a running example. Its two mutually recursive relations, \textit{ReachableMethod} and \textit{CallEdge}, exercise every rewrite of the pipeline. Figure~\ref{fig:midlog-cha} shows the Midlog\xspace\ snapshots after the second and third phases, and Figure~\ref{fig:cgcha-multihead} the Midlog\xspace\ form after the multi-head optimization of Section~\ref{sec:opt-multihead}, together with the recursive SQL produced from it.

\subsection{Datalog Normalization}\label{sec:normalization}

DLSQL\xspace accepts the full surface syntax of Soufflé-style Datalog (multi-head rules, body disjunctions, type aliases, repeated head variables, and constants or arithmetic expressions inside atoms) and rewrites it into textbook Datalog: every rule a single-headed Horn clause, every head a tuple of distinct variables, every atom argument a variable~\cite{abiteboul1995foundations}:

\begin{enumerate}[leftmargin=*]
  \item \emph{Disjunction split}: a body with $L_1; L_2$ splits into two rules, repeated until disjunction-free.
  \item \emph{Distinct head variables}: $R(x,x)$ becomes $R(x, x_2)$ with $x_2 = x$ added to the body.
  \item \emph{Atom variable-only}: arguments that are constants or arithmetic expressions are lifted out into fresh variables bound by equalities, e.g., $R(1, y)$ becomes $R(v, y)$ with $v = 1$.
  \item \emph{Existential elimination}: every $\_$ wildcard is given a fresh, otherwise-unused name.
  \item \emph{Dead-rule elimination}: rules unreachable from \texttt{.output} relations are removed by a backward reachability analysis.
\end{enumerate}

\noindent The result is a canonical \texttt{DatalogProgram} in the shape stated above, with every constant or expression exposed as a comparison atom. This shape enables the uniform per-\textsc{SCC}\xspace construction of the next phase and the direct mapping to a \sqlk{SELECT}--\sqlk{FROM}--\sqlk{WHERE} skeleton at the end of the pipeline.

For the \texttt{cgcha} example of Figure~\ref{fig:cha}, existential elimination rewrites the rule \textit{ReachableMethod}(\textit{ce}) \;\mbox{:-}\; \textit{CallEdge}(\_, \textit{ce}) into \textit{ReachableMethod}(\textit{ce}) \;\mbox{:-}\; \textit{CallEdge}(\texttt{a1}, \textit{ce}), with \texttt{a1} carrying through to Figure~\ref{fig:midlog-cha}.

\subsection{Datalog to Midlog\xspace}\label{sec:dl-to-midlog}

This phase translates the canonical program into Midlog\xspace{} with top-level rules in predicate dependency order, in two steps: an \textsc{SCC}\xspace{} analysis and a per-\textsc{SCC}\xspace{} construction. The transformation pass that turns each \textsc{SCC}\xspace{} into a single-head form is deferred to Midlog\xspace normalization (Section~\ref{sec:midlog-normalization}).

\noindent\textbf{Step 1: \textsc{SCC}\xspace analysis and topological ordering.} Tarjan's linear-time algorithm~\cite{tarjan1972depth} computes the \textsc{SCC}\xspace{}s of the predicate dependency graph (Section~\ref{sec:background}) in topological order, which becomes the order of Midlog\xspace{} rules and ultimately of \textsc{CTE}\xspace{}s, so each rule references only relations already defined and the heads of its own \textsc{SCC}\xspace.

\noindent\textbf{Step 2: per-\textsc{SCC}\xspace construction.} For each \textsc{SCC}\xspace, we build one top-level rule. A singleton \textsc{SCC}\xspace becomes a $\mathit{Def}$ whose body is the \sqlk{union} of all conjunctions defining its head (invariant I2). A multi-member \textsc{SCC}\xspace becomes a transient $\mathit{MultiDef}$ with one branch per relation, built the same way. Each conjunction is canonicalized so the head's declared columns appear positionally as body variables: for a head $R(x_1, \ldots, x_k)$ with columns $(c_1, \ldots, c_k)$, every body occurrence of $x_i$ is renamed to $c_i$ (invariant I3), after $\alpha$-renaming colliding body variables; projection through the head is then implicit. Column names come from \texttt{.decl} when present, otherwise from the first defining rule's head (invariant I4).

For \texttt{cgcha}, \textsc{SCC}\xspace analysis groups the mutually-recursive \textit{ReachableMethod} and \textit{CallEdge} into the multi-member \textsc{SCC}\xspace of Figure~\ref{fig:midlog-cha-initial}. The per-\textsc{SCC}\xspace{} construction packages it as a $\mathit{MultiDef}$ with two branches, one per relation, each branch's body the \sqlk{union} of that relation's two source rules.
This phase establishes invariants I1--I4; multi-member \textsc{SCC}\xspace{}s remain as $\mathit{MultiDef}$s, which the next phase eliminates.

\subsection{Midlog\xspace Normalization}\label{sec:midlog-normalization}

Midlog\xspace normalization eliminates every $\mathit{MultiDef}$, leaving each top-level rule a single-head $\mathit{Def}$. Single-head $\mathit{Def}$s pass through unchanged; only multi-member \textsc{SCC}\xspace{}s are touched.

\noindent\textbf{Multi-head elimination.} Consider a $\mathit{MultiDef}$ rule with branches $H_i(\vec{c_i}) \;\mbox{:-}\; B_i$, where each body is the \sqlk{union} of conjunctions. We introduce a fresh synthetic predicate, named $H_1\_\!\cdots\!\_H_n$,  concatenating the names $H_i$ that carries an extra \emph{tag} column recording a tuple's source branch, together with the column tuple $\vec{c}$ whose schema subsumes every $H_i$. Our implementation assumes that all columns of an \textsc{SCC}\xspace share the same type $\tau$; in all our benchmarks, $\tau$=\sqlk{number}.
Thus, $\vec{c} = (c_1,\dots,c_a)$ with $a$ the largest
arity; for example, $R(x,y)$ and $S(z,u,v)$ yield $R\_S(\mathit{tag}, c_1, c_2, c_3)$. The combined predicate becomes the \emph{primary} $\mathit{Def}$ of the recursive group; each original head is reintroduced as a local projection inside the primary and a hoisted top-level projection outside it:

\[
\begin{array}{l}
H_1\_\!\cdots\!\_H_n(\mathit{tag}, \vec{c}) \;\mbox{:-}\; \\
\quad \sqlk{local}\;  H_1(\vec{c_1}) \;\mbox{:-}\; H_1\_\!\cdots\!\_H_n(T_1, \vec{p_1}). \\
\quad \cdots \\
\quad \sqlk{local}\;  H_n(\vec{c_n}) \;\mbox{:-}\; H_1\_\!\cdots\!\_H_n(T_n, \vec{p_n}). \\
\quad B_1[\vec{q_1}=\sqlk{NULL}, \,\mathit{tag} = T_1\,] \;\sqlk{union}\; 
\\
\quad 
\cdots  
\;\sqlk{union}\; 
\\
\quad 
B_n[\vec{q_n}=\sqlk{NULL}, \,\mathit{tag} = T_n\,]. \\
H_1(\vec{c_1}) \;\mbox{:-}\; H_1\_\!\cdots\!\_H_n(T_1, \vec{p_1}). \\
\cdots \\
H_n(\vec{c_n}) \;\mbox{:-}\; H_1\_\!\cdots\!\_H_n(T_n, \vec{p_n}).
\end{array}
\]

The locals are visible only inside the primary's body. Each local and hoisted projection reads the primary through $\vec{p_i}$: the
tuple $\vec{c}$ with the slots assigned to $H_i$ renamed to $\vec{c_i}$ and the
remaining slots $\vec{q_i}$ bound to fresh variables. For $R$ above,
the local body has the columns $(x, y, c_3)$. The combined-relation body unions the branch bodies, each with the equalities for \sqlk{NULL} padding missing columns $\vec{q_i}$, and $\mathit{tag} = T_i$ for distinct constants $T_i$. Every body literal that referenced some $H_j$ of the \textsc{SCC}\xspace now resolves to the corresponding local. Thus, the entire group becomes a single self-recursion on $H_1\_\!\cdots\!\_H_n$. The trailing top-level projections expose each $H_i$ to rules outside the \textsc{SCC}\xspace and to \texttt{.output} declarations (invariant I6).

Applied to the \texttt{cgcha} \textsc{SCC}\xspace $\{\textit{ReachableMethod}, \textit{CallEdge}\}$, this rewrite introduces the combined primary, tags its tuples \texttt{RM} or \texttt{CE}, and reintroduces both relations as projection locals and hoisted siblings (Figure~\ref{fig:midlog-cha-normalized}).

\noindent\textbf{The tagged relation.} The combined primary has the schema $(\mathit{tag}{:}\,\mathit{int},
c_1{:}\,\tau,\dots,c_a{:}\,\tau)$; the $j$-th column of $H_i$ occupies $c_j$, and a
head with $a_i<j$ pads $c_j$ with \sqlk{NULL} 
through the equality
$c_j = \sqlk{NULL}$. If columns of an SCC have mixed types, the same construction
applies per type, allocating for each type the largest number of columns of that
type among the heads.
Note that a head column bound only by an equality comparison (e.g., $c_2 = \sqlk{NULL}$ and $\mathit{tag} = \texttt{RM}$ in Figure~\ref{fig:midlog-cha-normalized}) remains projectable thanks to I3{}:
the SQL emitter places the compared constant directly in the \sqlk{SELECT} list (e.g., \sqlk{SELECT} \texttt{RM}, $c_1$, \sqlk{NULL}).

\noindent\textbf{Invariants established.} This establishes invariants I5{} and I6{}; every top-level rule is now a $\mathit{Def}$ that is at most self-recursive.
Section~\ref{sec:opt-multihead} revisits this rewrite as an optimization: when a branch of a $\mathit{MultiDef}$ has a base case along with further restrictions on the predicate dependency graph, the synthetic combined relation can be avoided, and that branch is lifted to primary directly.

\noindent\textbf{Complexity.} The Datalog-to-Midlog\xspace translation is linear in the size of the program: it runs Tarjan's \textsc{SCC}\xspace algorithm on the predicate dependency graph and applies local per-\textsc{SCC}\xspace rewrites.
Both rewrites preserve data complexity, since the tagged fixpoint contains exactly one tuple per tuple of the original relations. However, the tagged encoding carries constant-factor overheads (the tag column, NULL padding to the widest head, and a $k$-way union per iteration) that peak on large \textsc{SCC}\xspace{}s of heterogeneous arities. Section~\ref{sec:exp:opt} provides an ablation study by varying the number of \textsc{IDB}\xspace{}s.

\subsection{Midlog\xspace to SQL}\label{sec:midlog-to-sql}

With the structural shape of the output already established, the translation to SQL is a near-mechanical walk.

\noindent\textbf{Top-level rules to \textsc{CTE}\xspace{}s.} Each top-level $\mathit{Def}$ becomes one \sqlk{WITH} entry, marked recursive iff its body or any of its (transitively nested) local's bodies, references the rule's own head. Thanks to Midlog\xspace{} normalization, no $\mathit{MultiDef}$ survives, so every \textsc{CTE}\xspace is single-headed.

\noindent\textbf{Base/step split for primary rules with locals.}  When a $\mathit{Def}$ has no locals, its body is emitted as the \sqlk{UNION} of its conjunctions, with \sqlk{SELECT DISTINCT} on each non-recursive branch to preserve Datalog's set semantics (recursive \textsc{CTE}\xspace{}s already deduplicate at the \sqlk{UNION} boundary). A $\mathit{Def}$ with locals is partitioned into \emph{base} branches (no reference to the recursive head) and \emph{step} branches. The former form one side of the outer \sqlk{UNION}; the latter are wrapped in an inner \sqlk{WITH} defining each local as a non-recursive \textsc{CTE}\xspace. The emitted \textsc{CTE}\xspace for a primary $P$ with base branches $B$, step branches $S$, and locals $\mathrm{Aux}$ has the shape
\[
P(\vec{c}) \;\sqlk{AS}\; (\;\mathrm{tr}(B) \;\sqlk{UNION}\; \sqlk{WITH}\;\mathrm{Aux}\;\;\mathrm{tr}(S)\;).
\]
Here $\mathrm{tr}(\cdot)$ is the \sqlk{SELECT} translation described next; each $A \in \mathrm{Aux}$ is visible only inside $P$'s recursive step, while the hoisted top-level projections (Section~\ref{sec:midlog-normalization}) emit non-recursive outer \textsc{CTE}\xspace{}s referencing the primary.

\noindent\textbf{Conjunction to \sqlk{SELECT}.} A Midlog\xspace conjunction is translated to one \texttt{SQLSingleStmt} (\sqlk{SELECT}--\sqlk{FROM}--\sqlk{WHERE}). The translator maintains an environment $\Gamma : \mathrm{Var} \to \mathrm{SQLTerm}$ mapping each Datalog variable to the column reference that first binds it. Every positive atom $R(\vec{t})$ adds a fresh table alias $t_k$ to the \sqlk{FROM} clause, and its argument list is walked:
\begin{itemize}[leftmargin=*]
  \item if $t_i$ is a fresh variable $v$ and $v \notin \Gamma$, bind $\Gamma(v) := t_k.c_i$ where $c_i$ is the corresponding declared column;
  \item if $t_i$ is a variable $v \in \Gamma$, emit the equality $\Gamma(v) = t_k.c_i$ as an implicit join condition;
  \item if $t_i$ is a constant or an arithmetic expression, emit the corresponding equality directly.
\end{itemize}
Comparison atoms become \sqlk{WHERE} predicates after substituting $\Gamma$; an equality binding an unbound variable is recorded by aliasing in $\Gamma$ rather than as a predicate, and a variable occurring only once (a former wildcard) binds in $\Gamma$ but adds no join condition. A negated atom $\neg R(\vec{t})$ becomes a correlated \sqlk{NOT EXISTS} subquery over a fresh alias of $R$, with one equality per already-bound variable and one per constant. The \sqlk{SELECT} clause is built last: by I3, each head column $c_i$ is already bound in $\Gamma$, so it simply projects $\Gamma(c_i)$ aliased to $c_i$.

\begin{figure*}[t]
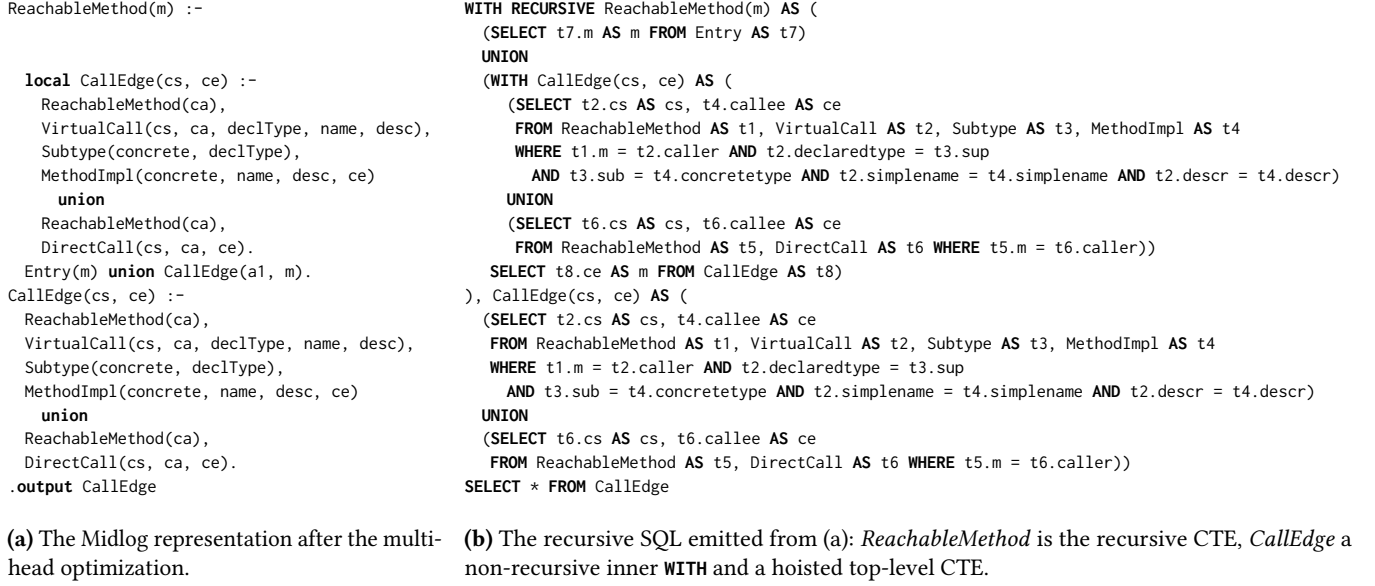

\begin{minipage}[t]{.32\textwidth}
\vspace{0pt}
\begin{lstlisting}[language=datalog,basicstyle=\ttfamily\scriptsize]
ReachableMethod(m) :-


  local CallEdge(cs, ce) :-    
    ReachableMethod(ca),
    VirtualCall(cs, ca, declType, name, desc),
    Subtype(concrete, declType),
    MethodImpl(concrete, name, desc, ce)
      union
    ReachableMethod(ca), 
    DirectCall(cs, ca, ce).
  Entry(m)  union  CallEdge(a1, m).
CallEdge(cs, ce) :-
  ReachableMethod(ca),
  VirtualCall(cs, ca, declType, name, desc),
  Subtype(concrete, declType),
  MethodImpl(concrete, name, desc, ce)
    union
  ReachableMethod(ca), 
  DirectCall(cs, ca, ce).
.output CallEdge
\end{lstlisting}
\subcaption{The Midlog\xspace representation after the multi-head optimization.}
\label{fig:midlog-cha-opt}
\end{minipage}\hfill
\begin{minipage}[t]{0.66\textwidth}
\vspace{0pt}
\begin{lstlisting}[basicstyle=\ttfamily\scriptsize,language=SQL,morekeywords={WITH}]
WITH RECURSIVE ReachableMethod(m) AS (
  (SELECT t7.m AS m FROM Entry AS t7)
  UNION
  (WITH CallEdge(cs, ce) AS (
     (SELECT t2.cs AS cs, t4.callee AS ce
      FROM ReachableMethod AS t1, VirtualCall AS t2, Subtype AS t3, MethodImpl AS t4
      WHERE t1.m = t2.caller AND t2.declaredtype = t3.sup
        AND t3.sub = t4.concretetype AND t2.simplename = t4.simplename AND t2.descr = t4.descr)
     UNION
     (SELECT t6.cs AS cs, t6.callee AS ce
      FROM ReachableMethod AS t5, DirectCall AS t6 WHERE t5.m = t6.caller))
   SELECT t8.ce AS m FROM CallEdge AS t8)
), CallEdge(cs, ce) AS (
  (SELECT t2.cs AS cs, t4.callee AS ce
   FROM ReachableMethod AS t1, VirtualCall AS t2, Subtype AS t3, MethodImpl AS t4
   WHERE t1.m = t2.caller AND t2.declaredtype = t3.sup
     AND t3.sub = t4.concretetype AND t2.simplename = t4.simplename AND t2.descr = t4.descr)
  UNION
  (SELECT t6.cs AS cs, t6.callee AS ce
   FROM ReachableMethod AS t5, DirectCall AS t6 WHERE t5.m = t6.caller))
SELECT * FROM CallEdge
\end{lstlisting}
\subcaption{The recursive SQL emitted from (a): \textit{ReachableMethod} is the recursive \textsc{CTE}\xspace, \textit{CallEdge} a non-recursive inner \sqlk{WITH} and a hoisted top-level \textsc{CTE}\xspace.}
\label{fig:cgcha-sql}
\end{minipage}

\vspace{-0.3cm}
\caption{The Midlog\xspace representation and the recursive SQL for \texttt{cgcha} with the multi-head optimization. No combined relation is introduced; instead, \textit{ReachableMethod} is lifted to primary, with \textit{CallEdge} nested as a local and hoisted as a top-level sibling.}
\label{fig:cgcha-multihead}
\vspace{-0.2cm}
\end{figure*}

For \texttt{cgcha}, the base/step split shapes the recursive SQL directly: \textit{ReachableMethod}'s \textsc{CTE}\xspace{} unions a base branch over \textit{Entry} with a step branch carrying its companion \textit{CallEdge} as an inner non-recursive \sqlk{WITH} (\textit{CallEdge} is also hoisted as a top-level \textsc{CTE}\xspace); the virtual-call resolution over \textit{VirtualCall}, \textit{Subtype}, and \textit{MethodImpl} becomes implicit joins in the \sqlk{WHERE} clause, which the database engine optimizes like any non-recursive join query. Figure~\ref{fig:cgcha-sql} shows the recursive SQL produced for \texttt{cgcha} with the multi-head optimization of Section~\ref{sec:opt-multihead}, side by side with the Midlog\xspace\ form (Figure~\ref{fig:midlog-cha-opt}) it is emitted from.

\subsection{Final Statement and Output}\label{sec:final-output}

After all top-level \textsc{CTE}\xspace{}s are emitted, the compiler emits a trailing \sqlk{SELECT * FROM} the relation declared with \texttt{.output}. A separate companion file emits \sqlk{CREATE TABLE} statements for each \texttt{.input}-declared \textsc{EDB}\xspace, with column types translated from the Datalog schema. For \texttt{cgcha}, the \texttt{.output} relation is \textit{CallEdge}, so the statement of Figure~\ref{fig:cgcha-sql} closes with \sqlk{SELECT * FROM} \textit{CallEdge}, mirroring the \texttt{.output} directive carried through Midlog\xspace\ (Figure~\ref{fig:midlog-cha-opt}), and the companion file creates its five \textsc{EDB}\xspace tables.

The unparser renders the SQL IR in the target dialect. Identifiers colliding with reserved keywords (\texttt{from}, \texttt{to}, \texttt{select}, \ldots) are quoted, and \sqlk{UNION} lists with $n>2$ branches are rendered as nested binary unions. DuckDB, Umbra, and Hyper accept the generated code unmodified; Section~\ref{sec:exp:portability} presents the dialect matrix for four further engines and the rewrites that close the gap.

\section{Optimizations}\label{sec:opt}

The four-phase pipeline of Section~\ref{sec:compiler} translates Linear Datalog\xspace programs into well-formed recursive SQL; two further passes optimize the result.
The first avoids the synthetic combined relation that Midlog\xspace normalization introduces for multi-member \textsc{SCC}\xspace{}s; the second discovers functional dependencies over \textsc{EDB}\xspace{}s and exposes them as primary keys.

\subsection{Multi-Head Optimization}\label{sec:opt-multihead}

The tag rewrite of Midlog\xspace normalization (Section~\ref{sec:midlog-normalization}) is universal but pays a price: a $\mathit{tag}$ column, tuples widened to accommodate every branch, and one recursive \textsc{CTE}\xspace{} unioning all branch bodies. When the \textsc{SCC}\xspace admits a natural primary, this overhead is avoidable.

\noindent\textbf{Rewrite condition.} The optimization is applicable to a $\mathit{MultiDef}$ when two conditions hold. First, exactly one branch $H_p$ is \emph{self-recursive} (references its own head) and has a base conjunction free of \textsc{SCC}\xspace literals. Second, $H_p$ is a \emph{feedback vertex} of the \textsc{SCC}\xspace's predicate dependency graph: removing it leaves the remaining branches, the \emph{companions}, acyclic. The pass lifts $H_p$ to the top-level $\mathit{Def}$ and attaches the companions as non-recursive locals in topological order. This removes the need for the $\mathit{tag}$ column and the combined tagged relation. The companions that are exposed (referenced from a later \textsc{SCC}\xspace\ or declared as an output relation) are hoisted as top-level siblings (invariant I6).

\noindent\textbf{Fallback.} If no branch is self-recursive, any branch with a base conjunction may serve as primary, subject to the same feedback-vertex condition. 
When no candidate is a feedback vertex, the compiler falls back to the tag rewrite.\footnote{Note that a generalized version of the multi-head optimization is to find the minimum feedback vertex set, which is an NP-hard problem~\cite{DBLP:books/fm/GareyJ79}. Our algorithm is linear and only finds a feedback vertex set of size one, whereas the generalized one can decrease the number of tagged IDBs.}
For instance, in $A \;\mbox{:-}\; \mathit{X}$; $A \;\mbox{:-}\; A, \mathit{Y}$; $A \;\mbox{:-}\; B$; $B \;\mbox{:-}\; C$; $C \;\mbox{:-}\; B$; $C \;\mbox{:-}\; A$, removing the only self-recursive branch $A$ leaves $B$ and $C$ mutually recursive, so the pass falls back to the tag rewrite.

For \texttt{cgcha}, the normalized Midlog\xspace{} (Figure~\ref{fig:midlog-cha-normalized}) fuses the \textsc{SCC}\xspace $\{\textit{ReachableMethod}, \textit{CallEdge}\}$ into a tagged combined relation. The optimization instead selects \textit{ReachableMethod} as the primary and nests \textit{CallEdge} as a local, hoisted as a top-level sibling (Figure~\ref{fig:midlog-cha-opt}); the resulting SQL (Figure~\ref{fig:cgcha-sql}) ranges over the original single-column \textit{ReachableMethod}. This is the path taken on every benchmark; the tagged fallback remains for programs that fail the condition, and Section~\ref{sec:exp:opt} measures its cost on all three backends.

\noindent\textbf{Semantic Equivalence.} Both normalized forms of Midlog\xspace{} preserve the source program's semantics in the sense of Section~\ref{sec:midlog}. The tagged form is semantics-preserving on the entire fragment: its fixed point is the tag-disjoint union of the \textsc{SCC}\xspace's relations, with one tagged tuple per original tuple. The optimized form is semantics-preserving exactly under the feedback-vertex condition: an acyclic companion graph makes every companion a non-recursive \textsc{IDB}\xspace{} over the primary relation and the \textsc{EDB}\xspace{}s, whose topological evaluation reproduces the original \textsc{SCC}\xspace's immediate-consequence step.

\noindent\textbf{Local-\textsc{IDB}\xspace inlining.} The normalized Midlog\xspace{} can introduce local \textsc{IDB}\xspace{}s. In particular, multi-head optimization nests each companion as a non-recursive \sqlk{WITH} inside the recursive term (\textit{CallEdge} in Figure~\ref{fig:cgcha-sql}), a syntax that stricter engines reject. A further Midlog\xspace-to-Midlog\xspace pass inlines each local \textsc{IDB}\xspace's rules into the recursive term, removing every nested \sqlk{WITH}. Section~\ref{sec:exp:portability} evaluates the portability this rewrite enables.

\subsection{Functional-Dependency Discovery for \textsc{EDB}\xspace{}s}\label{sec:opt-fd}

Much of a relational engine's optimization power rests on functional dependencies and keys: they let the planner turn a join into a single-tuple index probe and pick the smaller build side. Soufflé's \texttt{.decl} signatures declare no such constraints, so the engine treats every \textsc{EDB}\xspace table as an arbitrary relation. DLSQL\xspace\ recovers the missing constraints from the input data and passes them to the engine as schema-level keys.

\noindent\textbf{Discovery.} Before emitting the \sqlk{CREATE TABLE} script of Section~\ref{sec:final-output}, the compiler runs an FD-discovery pass~\cite{huhtala1999tane,papenbrock2016hybrid} over the supplied \textsc{EDB}\xspace tuples, enumerating minimal non-trivial FDs and minimal candidate keys per relation. Discovery operates on the materialized input, so the FDs are exact for the supplied workload, not schema-level guarantees. The schema script is regenerated per dataset; a stale key fails loudly at load time (the engine rejects the violating insertion), never silently at query time.

\noindent\textbf{Hints emitted to the engine.} For each \textsc{EDB}\xspace we emit a \sqlk{PRIMARY KEY} declaration on the smallest discovered candidate key whose columns are all referenced from at least one recursive \textsc{CTE}\xspace; recursive \textsc{CTE}\xspace{}s that join against keyed \textsc{EDB}\xspace relations then gain index-supported probes on the join's inner side, with no change to the generated query.

\noindent\textbf{Enforcement semantics.} The same schema script runs unmodified on every backend, yet the keys carry different semantics per engine. DuckDB and Umbra enforce the declared keys during data loading and honor them during recursive-\textsc{CTE}\xspace planning. Hyper accepts them only as \sqlk{ASSUMED} constraints, pure optimizer hints, but not checked by the engine~\cite{hyper-createtable}; there, the per-dataset regeneration above carries the correctness burden alone.

\noindent\textbf{Cost.} FD discovery is worst-case exponential in the number of attributes~\cite{huhtala1999tane,papenbrock2016hybrid,DBLP:conf/icdt/Nakos0T25}. In our case, this is not problematic as our \textsc{EDB}\xspace arities are small; discovery is a one-time offline pass per dataset, and key construction happens during data loading. Section~\ref{sec:exp:opt} shows where key construction can improve the performance.

\section{Experimental Results}\label{sec:exp}

We evaluate DLSQL\xspace on five canonical program-analysis workloads drawn from prior Datalog benchmarks, aiming to answer the following five questions:

\begin{itemize}[topsep=2pt,itemsep=1pt,leftmargin=*]
\item How does DLSQL\xspace compare end-to-end against Soufflé, FlowLog, Logica, and RecStep (cf. Section~\ref{sec:exp:e2e})?
\item How much do the two optimizations of Section~\ref{sec:opt} contribute (cf. Section~\ref{sec:exp:opt})?
\item How does the engine's ``free'' parallelism compare with Datalog systems' hand-built parallelism, and what are the root causes when it falls behind (cf. Section~\ref{sec:exp:parallel})?
\item How portable is the generated SQL across relational engines (cf. Section~\ref{sec:exp:portability})?
\item How does performance scale with input size and recursion depth (cf. Section~\ref{sec:exp:scaling})?
\end{itemize}

\subsection{Experimental Setup}\label{sec:exp:setup}

We compare DLSQL\xspace against Soufflé~\cite{souffle}, the dominant Datalog engine using C++ code generation (compiled mode, 1.5--2$\times$ faster than its interpreter~\cite{flowlog}), and FlowLog~\cite{flowlog}, a recent Datalog compiler lowering to Differential Dataflow in Rust. We further compare against two relational-backend systems: Logica~\cite{logica}, compiling Datalog to SQL on DuckDB but driving the recursion from an external Python loop (we use its iterative mode), and RecStep~\cite{recstep2019}, coupling a modified QuickStep engine with an external semi-naive interpreter. We could not compare with Flan~\cite{flan}, as it has no public artifact. DLSQL\xspace compiles each Linear Datalog\xspace program once to a single \sqlk{WITH RECURSIVE} block, run on three RDBMS: DuckDB~\cite{duckdb} (in-process, columnar), Umbra~\cite{neumann2020umbra} (in-memory, JIT-compiling~\cite{DBLP:conf/sigmod/ShaikhhaKPBD016,DBLP:journals/pvldb/Neumann11}), and Hyper~\cite{kemper2011hyper} (Umbra and Hyper are closed-source but binaries are freely available).
All systems run at pinned versions (DuckDB 1.5.5, Umbra \texttt{umbradb/umbra:26.06}, Hyper 0.0.24457, Soufflé 2.4, FlowLog \texttt{1c3be66}, RecStep \texttt{b7b41b7} with QuickStep-Datalog \texttt{ef3e350}, Logica 1.3.1415926535897). We set only the thread count, with all remaining settings being engine defaults.

We run 5 timed repeats and report the median $\pm \sigma$ (Tables~\ref{tab:e2e} and~\ref{tab:portability}, and error bars in Figures~\ref{fig:ablation} and~\ref{fig:scaling}). Every configuration runs with a 15-minute timeout; configurations exceeding it are reported as DNF. We only measure the query execution time and exclude loading and compilation for every system (Soufflé: profiler-reported runtime minus per-relation loads; SQL engines: the query alone; Logica: the fixpoint loop alone, on pre-loaded connections). FlowLog's self-reported timer includes \textsc{EDB}\xspace loading (up to 44\%, \texttt{escape} at 8 threads); we subtract it using its own timestamps. RecStep's reported time is similarly load-corrected using its interpreter's timestamps. We validate every completed configuration's output cardinalities (not full tuple sets) against Soufflé.

Compilation costs are one-time and excluded: Soufflé's synthesized C++ takes 13.5--25.8\,s per program, whereas DLSQL\xspace is source-to-source with no native-compilation step. Inputs follow the Soufflé fact-file convention (one CSV per \textsc{EDB}\xspace), loaded into tables whose schemas DLSQL\xspace emits.

Each backend additionally has a \emph{+FD} configuration whose \sqlk{CREATE TABLE} schema declares the keys DLSQL\xspace discovers (Section~\ref{sec:opt-fd}), exactly what a hand-written SQL application would declare; the query is unchanged.

All experiments run on one machine (Ubuntu 22.04, 10-core Intel Xeon Silver 4210 @ 2.2\,GHz, 250\,GB RAM). We measure with threads=1 to isolate algorithmic cost and threads=8 to expose intra-query parallelism.

Table~\ref{tab:bench-summary} summarizes the five programs (a singleton predicate counts as its own \textsc{SCC}\xspace); they are the analyses of Section~\ref{sec:analyses}, with the three Java analyses running on facts from \textsc{antlr}~\cite{DBLP:conf/oopsla/BlackburnGHKMBDFFGHHJLMPSVDW06}, \texttt{csda} on a Linux-kernel value-flow graph~\cite{graspan}, and \texttt{borrow} on the Polonius borrow-checker facts~\cite{polonius}.
The inputs are considerably large; \texttt{escape} consumes the 42M-tuple \textit{VarPointsTo} relation as an \textsc{EDB}\xspace, \texttt{csda}'s \textit{Edge} relation has roughly 43M tuples, and \texttt{borrow}'s control-flow graph \textit{cfg\_edge} has around 48K tuples.

The real datasets fix one input per analysis. We complement them with a synthetic benchmark varying size and depth: a two-rule linear transitive closure over $\mathit{chains}(W,L)$ ($W$ disjoint chains of length $L$). As input, this benchmark has $W \cdot L$ edges, and produces $W(L{+}1)$ output tuples in $L$ semi-naive rounds.

\begin{table}[t]
\centering
\caption{Summary of the five program-analysis benchmarks, with input/output cardinalities and round counts. For \texttt{escape} and \texttt{borrow} with multiple \textsc{SCC}\xspace{}s, output and round counts are reported for the most time-consuming \textsc{SCC}\xspace{}.}
\label{tab:bench-summary}
\begin{footnotesize}
\setlength{\tabcolsep}{3.5pt}
\begin{tabular}{lrrrrrrr}
\hline
\textbf{Benchmark} & \rotatebox{90}{\textbf{\#EDBs}} & \rotatebox{90}{\textbf{\#IDBs}} & \rotatebox{90}{\textbf{\#SCCs}} & \rotatebox{90}{\textbf{\#Rules}} & \rotatebox{90}{\textbf{\#Input}} & \rotatebox{90}{\textbf{\#Output}} & \rotatebox{90}{\textbf{\#Rnds}} \\
\hline
\texttt{varpointsto} & 23 & 2 & 1 & 10 & 27.2M & 42.3M & 43 \\
\texttt{cgcha}       &  5 & 2 & 1 &  4 & 615K & 1.37M & 63 \\
\texttt{escape}      &  9 & 5 & 5 & 11 & 44.3M & 1{,}737 & 4 \\
\texttt{csda}        &  2 & 1 & 1 &  2 & 44.0M & 55.8M & 778 \\
\texttt{borrow}      &  4 & 5 & 5 &  9 & 71.8K & 292M & 1008 \\
\hline
\end{tabular}
\vspace{-0.4cm}
\end{footnotesize}
\end{table}

\subsection{End-to-End Results}\label{sec:exp:e2e}

\begin{table*}[t]
\centering
\caption{Query evaluation times at 8 threads (seconds; median $\pm \sigma$ of five repeats; speedup of medians vs.\ Soufflé; +FD for the DLSQL\xspace backends). DNF: exceeds the 900\,s timeout or fails during data loading.}
\label{tab:e2e}
\vspace{-0.4cm}

{\begin{footnotesize}\setlength{\tabcolsep}{2pt}
\begin{tabular}{lrrrrrrr}
\hline
& \textbf{Souffl\'e} & \textbf{FlowLog} & \textbf{Logica} & \textbf{RecStep} & \multicolumn{3}{c}{\textbf{DLSQL\xspace}} \\
\cline{6-8}
\textbf{Benchmark} & & & & & \textbf{DuckDB} & \textbf{Umbra} & \textbf{Hyper} \\
\hline
escape       & 1.14$\pm$0.01 & 6.68$\pm$0.02 (0.17$\times$) & 84.04$\pm$0.23 (0.01$\times$) & 21.96$\pm$0.95 (0.05$\times$) & 4.84$\pm$0.12 (0.24$\times$) & \textbf{0.32$\pm$0.09 (3.57$\times$)} & 4.18$\pm$0.40 (0.27$\times$) \\
cgcha        & 0.74$\pm$0.01 & 2.75$\pm$0.02 (0.27$\times$) & 29.66$\pm$0.11 (0.03$\times$) & 58.22$\pm$3.73 (0.01$\times$) & 1.45$\pm$0.16 (0.51$\times$) & \textbf{0.13$\pm$0.11 (5.59$\times$)} & 0.53$\pm$0.16 (1.40$\times$) \\
varpointsto  & 73.42$\pm$0.25 & 74.45$\pm$0.17 (0.99$\times$) & DNF & DNF & 46.70$\pm$0.44 (1.57$\times$) & \textbf{6.77$\pm$0.40 (10.84$\times$)} & 34.72$\pm$0.20 (2.11$\times$) \\
csda         & 35.01$\pm$0.10 & 25.02$\pm$0.09 (1.40$\times$) & DNF & DNF & 54.71$\pm$0.04 (0.64$\times$) & \textbf{22.41$\pm$0.01 (1.56$\times$)} & 53.34$\pm$0.17 (0.66$\times$) \\
borrow       & 217.54$\pm$0.58 & 318.08$\pm$0.83 (0.68$\times$) & DNF & DNF & 178.43$\pm$0.64 (1.22$\times$) & \textbf{15.20$\pm$0.09 (14.31$\times$)} & 102.11$\pm$0.26 (2.13$\times$) \\
\hline
Geomean      & 1.00$\times$ & 0.53$\times$ & -- & -- & 0.68$\times$ & \textbf{5.46$\times$} & 1.02$\times$ \\
\hline
\end{tabular}
\end{footnotesize}}
\vspace{-0.2cm}
\end{table*}

Table~\ref{tab:e2e} reports query evaluation times at 8 threads; the three DLSQL\xspace columns report the +FD configuration. Section~\ref{sec:exp:opt} isolates the optimizations, and Section~\ref{sec:exp:parallel} the thread count.

\noindent\textbf{The outcome depends on the RDBMS backend.} At 8 threads, DLSQL\xspace's geomean speedup over Soufflé is 5.46$\times$ on Umbra, 1.02$\times$ on Hyper, and 0.68$\times$ on DuckDB. DuckDB's three losses have two causes: on the small \texttt{escape} and \texttt{cgcha} there is too little work to amortize DuckDB's overheads (Soufflé finishes within ${\sim}1$\,s), and on \texttt{csda} the 778-round recursion pays DuckDB's per-round overhead (cf. Section~\ref{sec:exp:scaling}). At 1 thread, DuckDB becomes more competitive (geomean 0.96$\times$, winning three of five), so its 8-thread gap is due to parallelization, not algorithmic (cf. Section~\ref{sec:exp:parallel}). Hyper is competitive in the single-threaded setting as well (geomean 1.05$\times$). The order-of-magnitude wins belong to Umbra (2.42$\times$ already at 1 thread): at 8 threads it beats Soufflé on every benchmark, up to 14.31$\times$ on \texttt{borrow}, using only the emitted \sqlk{WITH RECURSIVE} query.

\noindent\textbf{FlowLog is consistently behind DLSQL\xspace.} FlowLog generates Differential Dataflow code. On all benchmarks, it is slower than Umbra, and except on \texttt{csda}, it is slower than DuckDB and Hyper. FlowLog overtakes Soufflé only on \texttt{csda} (1.40$\times$); its geomean is 0.53$\times$ of Soufflé.

\noindent\textbf{Logica isolates the design choice.} Logica targets the same DuckDB backend as DLSQL\xspace, so the same-engine comparison isolates the compilation design: at 8 threads Logica trails DLSQL\xspace by 17.4$\times$ on \texttt{escape} and 20.5$\times$ on \texttt{cgcha} (cf. Table~\ref{tab:e2e}; at 1 thread the gaps grow to 46.6$\times$ and 24.5$\times$), and it exceeds the timeout on the other three. The gap is the design: Logica drives each semi-naive iteration from an external Python loop, whereas DLSQL\xspace hands the engine one recursive query.

\noindent\textbf{RecStep.} RecStep completes only \texttt{escape} and \texttt{cgcha}, far behind DLSQL\xspace on every backend (cf. Table~\ref{tab:e2e}); on the rest it does not finish at any tried thread count, exceeding the 15-minute timeout or failing during data loading.

\subsection{\texorpdfstring{Impact of the Optimizations}{Impact of the Optimizations}}\label{sec:exp:opt}

\noindent\textbf{Multi-head optimization.} The multi-head optimization (Section~\ref{sec:opt-multihead}) removes the combined tagged relation Midlog\xspace uses for mutually recursive \textsc{SCC}\xspace{}s. In our benchmark, only \texttt{cgcha} and \texttt{varpointsto} contain a multi-member \textsc{SCC}\xspace (both two-member), so we run the real-data ablation on \texttt{cgcha} and consider larger \textsc{SCC}\xspace{}s synthetically.

\begin{figure}[t]
\centering
\includegraphics[width=\linewidth]{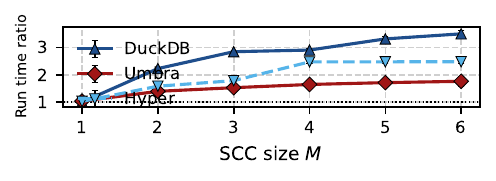}
\vspace{-1cm}
\caption{Overhead of the combined tagged relation vs.\ the multi-head optimization on synthetic $M$-member \textsc{SCC}\xspace{}s (chains, $W{=}100$K, $L{=}60$, 8 threads).}
\vspace{-0.4cm}
\label{fig:ablation}
\end{figure}

On \texttt{cgcha} at 8 threads, the combined tagged relation costs 1.45$\times$ on DuckDB (2.11\,s versus 1.46\,s), 1.84$\times$ on Umbra (0.233\,s versus 0.126\,s), and 2.02$\times$ on Hyper (1.11\,s versus 0.55\,s). Figure~\ref{fig:ablation} quantifies the synthetic case of growing \textsc{SCC}\xspace size $M$ on a family of $M$ mutually recursive relations. The tagged overhead grows with $M$, reaching 3.51$\times$ (DuckDB), 1.76$\times$ (Umbra), and 2.48$\times$ (Hyper) at $M{=}6$. The optimization removes this overhead whenever its rewrite condition holds.

\begin{figure}[t]
\centering
\includegraphics[width=\linewidth]{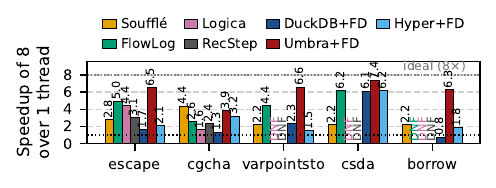}
\vspace{-1cm}
\caption{Each system's speedup at 8 threads over its own single-threaded setting. DNF: no completed 1-thread run (FlowLog \texttt{borrow}; Logica/RecStep beyond \texttt{escape}/\texttt{cgcha}).}
\label{fig:parallel}
\vspace{-0.4cm}
\end{figure}

\noindent\textbf{Functional dependencies.} We compare each backend with and without the \textsc{EDB}\xspace-side FD keys at 8 threads. The compiled query is identical, so the speedup isolates the key-based optimizations. The keys matter most on Umbra: 1.68$\times$ on \texttt{escape} and 1.49$\times$ on \texttt{csda} (a better hash-join build side), and at most 1.06$\times$ elsewhere. On DuckDB the effect is neutral, and Hyper, which accepts the keys only as unenforced \sqlk{ASSUMED} hints, shows at most 1.01$\times$.

The FD effect replicates beyond our three main backends (cf. Section~\ref{sec:exp:portability}): SQLite speeds up \texttt{escape} 2.2$\times$ and \texttt{csda} 1.15$\times$, and PostgreSQL \texttt{escape} 1.9$\times$ and \texttt{cgcha} 5.8$\times$. Keys can also hurt where the recursion is cheap: SQLite's and MySQL's \texttt{cgcha} regress slightly, as index maintenance is no longer free. FD discovery is a one-time offline pass per dataset, and key building happens in the untimed load phase, symmetric with every system's load exclusion; MySQL's key build alone exceeds 15 minutes on the 42M-row \textit{VarPointsTo} and 43M-row \textit{Edge}.

\subsection{Impact of Parallelization}\label{sec:exp:parallel}

Figure~\ref{fig:parallel} reports each engine's 8-thread speedup over its own single-threaded runtime on every benchmark.

\noindent\textbf{Umbra scales best, but not uniformly.} Umbra gains 6.3--7.4$\times$ on the four larger workloads but only 3.9$\times$ on \texttt{cgcha}; the gains are monotonic in thread count, thanks to morsel-driven parallel execution.

\noindent\textbf{DuckDB scales unevenly, and \texttt{borrow} anti-scales.} DuckDB matches Umbra only on \texttt{csda}; elsewhere it gains little, and on \texttt{borrow} more threads make it slower: 136.3\,s at 1 thread vs.\ 178.4\,s at 8 (0.76$\times$).

To find the root cause, we decomposed \texttt{borrow} into its \textsc{SCC}\xspace{}s and re-ran each alone. The \textsc{SCC}\xspace behind the 1008-round recursive \textsc{CTE}\xspace is the only one that anti-scales on DuckDB (0.59$\times$ at 8 threads), while Umbra achieves 7.02$\times$ on the same \textsc{SCC}\xspace. Thus, the effect is engine-specific, not workload-inherent. \sqlk{EXPLAIN ANALYZE} reveals the cause: in each round, DuckDB launches a parallel scan of the static \textit{cfg\_edge} relation, rescanning its 48{,}801 rows nearly in full $\sim921$ times across the 1008 rounds. As the relation is small and the per-round delta is tiny, each round performs too little work to amortize the fixed overhead of DuckDB's parallel scan, and additional threads only add coordination cost, hence the anti-scaling. In contrast, on \texttt{csda} the static relation has $\sim43$M rows, so each round performs enough work to amortize this overhead, and the recursion parallelizes well.

\noindent\textbf{Hyper scales, but unevenly.} Hyper's 8-thread speedups range from 1.55$\times$ (\texttt{varpointsto}) to 6.20$\times$ (\texttt{csda}); the geomean thread speedup (2.6$\times$) is within 3\% of Soufflé's, so the parity of Section~\ref{sec:exp:e2e} holds at both thread counts.

\subsection{\texorpdfstring{Portability across Engines}{Portability across Engines}}\label{sec:exp:portability}

We ran the generated SQL on seven engines; Table~\ref{tab:portability} reports the four beyond our main backends. DuckDB, Umbra, and Hyper accept the queries exactly as generated for all five benchmarks. Similarly, PostgreSQL, MariaDB, MySQL, and SQLite accept \texttt{escape}, \texttt{csda}, and \texttt{borrow} as generated. The remaining two benchmarks nest local \textsc{IDB}\xspace{}s as inner \textsc{CTE}\xspace{}s, which MariaDB, MySQL, and SQLite reject; the local-\textsc{IDB}\xspace inlining pass (Section~\ref{sec:opt-multihead}) removes them, after which all three accept and validate all five (SQLite after de-parenthesizing \sqlk{UNION} branches; MariaDB/MySQL under \texttt{ANSI\_QUOTES}). 

PostgreSQL accepts the nesting but allows only a single self-reference globally (a relaxing patch~\cite{hirn2022multiref} was never merged), rejecting both even after inlining. A minimal \emph{\textsc{CTE}\xspace-shadowing} rewrite closes this gap: an inner \sqlk{WITH} re-declares the recursive name, leaving one counted self-reference (sound, as the working table is a fixed snapshot per evaluation of the recursive term). With the rewrite, PostgreSQL accepts both, validating both \texttt{cgcha} and \texttt{varpointsto}. Because the recursion is linear, a fully collapsed single-self-reference form exists for every program in our suite, validated manually on all 25 (program, engine) combinations on validation data. Portability does not imply performance: they mostly time out on the two largest workloads (cf. Table~\ref{tab:portability}).

\begin{table}[t]
\centering
\caption{Query evaluation time on the four engines beyond our main backends (seconds; median $\pm \sigma$; best completed configuration; DNF: exceeds 900\,s). \texttt{cgcha}/\texttt{varpointsto} need the inlining pass on MariaDB/MySQL/SQLite and the \textsc{CTE}\xspace-shadowing rewrite on PostgreSQL.}
\label{tab:portability}
\begin{footnotesize}
\setlength{\tabcolsep}{3.5pt}
\begin{tabular}{lrrrr}
\hline
\textbf{Benchmark} & \textbf{PostgreSQL} & \textbf{MariaDB} & \textbf{MySQL} & \textbf{SQLite} \\
\hline
\texttt{escape}      & 24.3 $\pm$ 7.1 & 816.0 $\pm$ 0.4 & 331.9 $\pm$ 0.4 & 54.5 $\pm$ 0.1 \\
\texttt{cgcha}       & 21.4 $\pm$ 5.6 & DNF             & 19.8 $\pm$ 0.0  & 12.2 $\pm$ 0.0 \\
\texttt{varpointsto} & DNF            & DNF             & DNF             & DNF \\
\texttt{csda}        & DNF            & DNF             & DNF             & 283.1 $\pm$ 0.6 \\
\texttt{borrow}      & 662 $\pm$ 129  & DNF             & DNF             & DNF \\
\hline
\end{tabular}
\vspace{-0.4cm}
\end{footnotesize}
\end{table}

\begin{figure}[t]
\centering
\includegraphics[width=\linewidth]{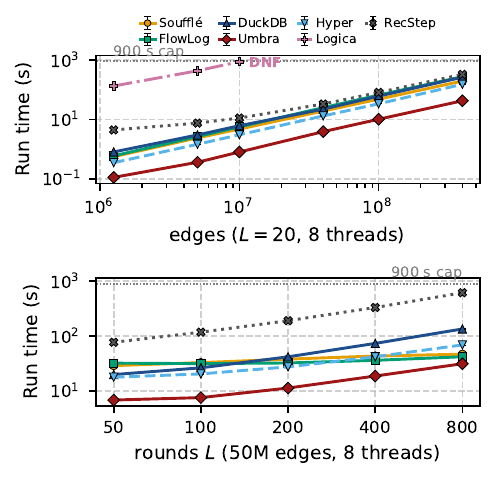}
\caption{Scaling at 8 threads: (a) run time vs.\ edge count at $L{=}20$; (b) run time vs.\ round count $L$ at 50M edges. Only Logica hits its limit: truncated in (a), absent from (b).}
\label{fig:scaling}
\end{figure}

\subsection{\texorpdfstring{Scaling with Input Size and Recursion Depth}{Scaling with Input Size and Recursion Depth}}\label{sec:exp:scaling}

The previous sections considered a single real dataset input per analysis. This section uses the linear transitive closure over $\mathit{chains}(W,L)$ (cf. Section~\ref{sec:exp:setup}). We run 11 configurations on all seven systems at 8 threads. First, we fixed the number of rounds ($L{=}20$) and varied the number of edges (1.25M--400M). Second, we fixed the number of edges and varied the number of round counts ($L{=}50$--$800$). We validated the correctness of all 69 completed cells out of 77 (the 8 missing cells are Logica's timeouts).

Figure~\ref{fig:scaling} shows the input scaling behavior. First, run time is near-linear in edge count for a fixed number of rounds. Out of six systems reaching 400M edges, five grow 82--118$\times$ for 80$\times$ more edges; RecStep is sublinear (44$\times$, fixed overhead amortizing as the number of edges grows). Second, the round count separates the execution models: raising $L$ from 50 to 800 at 50M edges leaves the Datalog-native systems (Soufflé and FlowLog) nearly flat (1.3--1.6$\times$), costs the relational engines 3.9--6.8$\times$, and RecStep (separate SQL batches per round) 8.0$\times$. Third, Logica shows significantly inferior performance, reaching the 15-minute timeout beyond 10M edges; all others complete every cell.

\section{Related Work}\label{sec:related}

\noindent\textbf{Datalog engines for program analysis.} Soufflé~\cite{souffle} compiles Datalog to specialized C++ and pioneered many of the rule-level optimizations
behind modern Datalog-based program analysis. Datalog compilation is an active topic in the compiler community. Soufflé's interpreter~\cite{hu2021souffle} trades off the overhead of its compiler~\cite{scholz2016fast} for a 1.5--2$\times$ execution overhead. Lattice-based fixed points~\cite{madsen2016flix}, data-parallel compilation~\cite{gilray2021compiling}, incremental whole-program analysis~\cite{szabo2021incremental}, macro-based Datalog in Rust~\cite{sahebolamri2022ascent}, and declarative C static checkers~\cite{dura2024clog} are among Datalog efforts. Flan~\cite{flan} pushes further with an expressive front end and a multi-stage-programming runtime; LogicBlox~\cite{aref2015design} and its successor Rel~\cite{aref2025rel} integrate Datalog into a commercial relational platform. FlowLog~\cite{flowlog} lowers rules to a per-rule relational IR over Differential Dataflow with explicit incremental-maintenance support, rather than emitting standard \sqlk{WITH RECURSIVE} blocks for off-the-shelf engines. Doop~\cite{bravenboer2009strictly} is the canonical Datalog-formulated points-to analysis and the source of our \texttt{varpointsto} benchmark.  Polonius~\cite{polonius} reformulates Rust's borrow checker as Datalog and is the source of our \texttt{borrow} benchmark.

\noindent\textbf{RDBMS-backed Datalog.} Several systems evaluate Datalog on a relational engine, differing in where the fixpoint runs. RecStep~\cite{recstep2019} drives a modified QuickStep engine from an external interpreter that issues non-recursive SQL per semi-naive iteration; its evaluation reports losing to Soufflé on the CSDA workload we share, partly due to this per-iteration overhead. Logica~\cite{logica} compiles a Datalog dialect to SQL but either unrolls user-level recursion to a bounded depth or iterates via an external Python driver, never emitting a user-level recursive \textsc{CTE}\xspace. The SQL baseline in the original Soufflé compiler paper~\cite{scholz2016fast} is similarly an SQLite-based semi-naive driver issuing one statement per iteration. DLSQL\xspace takes the opposite route: the entire fixpoint runs inside one standard recursive query, so the host engine's optimizer sees the whole recursive computation. Section~\ref{sec:exp:e2e} isolates this choice: on the identical DuckDB backend, DLSQL\xspace is 17.4$\times$ and 20.5$\times$ faster than Logica on the two benchmarks Logica completes.

\noindent\textbf{Recursive SQL and RDBMS support for fixed points.} SQL:1999 introduced recursive CTEs~\cite{sql1999,eisenberg1999sql}, restricted to linear recursion; database research has explored the magic-sets transformation as a query rewriter~\cite{bancilhon1985magic}, semi-naive evaluation~\cite{bancilhon1985naive,balbin1987generalization}, subsumption~\cite{kostler1995fixpoint,DBLP:journals/pacmmod/ShaikhhaSSN24}, and proposed extensions for recursive SQL~\cite{hirn2023fix, DBLP:conf/edbt/PassingTHLSGK017}. RaSQL~\cite{gu2019rasql} and BigDatalog~\cite{shkapsky2016big} push Datalog through Spark SQL on distributed engines. Also, there have been efforts to compile user-defined functions to recursive SQL~\cite{burghardt2022fp,10.1145/3448016.3457272}. They employ a label column to dispatch among mutually recursive functions. Our tagged relation uses a similar idea, but at each round the engine joins an entire delta relation against the \textsc{EDB}\xspace{}s, the tag records a tuple's source relation, and the multi-head optimization removes it whenever the \textsc{SCC}\xspace has a feedback-vertex primary. Language-Integrated Recursive Queries~\cite{herlihy2026lir} embed fixpoint queries as a Scala DSL lowering to recursive SQL, explicitly identifying the absence of mutual-recursion support in existing SQL implementations as a motivating restriction. We instead treat the engine as an unmodified backend and concentrate the front-end work in the compiler, so the same input runs on multiple engines.

\noindent\textbf{Datalog-to-SQL compilation.} Compiling Datalog to SQL has a long history~\cite{ceri1989you,abiteboul1995foundations}; the textbook recipe is to emit one \textsc{CTE}\xspace per \textsc{IDB}\xspace predicate and order them topologically. Outside \textsc{SCC}\xspace{}s, DLSQL\xspace emits exactly this recipe: every non-recursive or self-recursive \textsc{IDB}\xspace becomes one \textsc{CTE}\xspace, in topological order. Inside a multi-member \textsc{SCC}\xspace{}, the recipe stops working, since with the exception of MariaDB none of the engines we tested accepts mutually recursive \sqlk{WITH RECURSIVE} names; Midlog\xspace's tagged and multi-head rewrites resolve exactly this case, and Section~\ref{sec:exp:opt} measures their cost. DLSQL\xspace\ differs from prior work in its focus on (i) the Linear Datalog\xspace fragment as a portability target, validated by running the generated SQL on the seven engines of Section~\ref{sec:exp:portability}, and (ii) a dedicated intermediate language, Midlog\xspace, whose lexically-scoped local rules keep the output within that fragment for every supported program, by lifting a natural primary head when one exists or introducing a synthetic combined tagged relation. Raqlet~\cite{raqlet} provides the foundation for translating recursive query languages to each other, and CaQL~\cite{herlihy2025static} translates an embedded recursive query language to SQL. However, neither supports rewriting mutually recursive \textsc{SCC}\xspace{}s into the single-self-reference recursive queries that typical engines accept; Midlog\xspace's tagged and multi-head rewrites enable this.

\section{Conclusion}\label{sec:conclusion}

We presented DLSQL\xspace, a compiler from Linear Datalog\xspace to SQL, and showed that targeting modern RDBMS is often faster than state-of-the-art Datalog engines on canonical program analyses. Two ingredients underpin this result: the Linear Datalog\xspace fragment, rich enough to express call-graph construction, points-to analysis, escape analysis, dataflow analysis, and Polonius-style borrow checking, yet narrow enough to compile into a \sqlk{WITH RECURSIVE} block that runs across the seven relational engines (cf. Section~\ref{sec:exp:portability}); and the Midlog\xspace intermediate language, which absorbs the structural mismatch between Datalog and SQL (mutual recursion and different scoping rules) into a small set of rewrites, leaving SQL emission near-mechanical.
Two further passes refine the output: a multi-head optimization that lifts a natural primary head when one exists, and a functional-dependency discovery pass that exposes primary keys to the engine's optimizer.

Several directions remain. First, lifting the linearity restriction would broaden the source language to full Datalog, and its extensions, e.g., Datalog$^\circ$~\cite{DBLP:journals/jacm/KhamisNPSW24}. This can be achieved via three routes (non-standard SQL extensions for general fixpoints~\cite{hirn2023fix}, an external driver or unrolling in the style of RecStep~\cite{recstep2019} and Logica~\cite{logica}, or compile-time rewriting through non-linear-to-linear transformations~\cite{sagiv1988optimizing}). Second, \textsc{SCC}\xspace-level linearity detection could serve as a backend-selection pass inside existing Datalog engines, routing linear \textsc{SCC}\xspace{}s to a relational backend and evaluating the rest natively. Third, lattice-valued aggregations (as in Flix) are a natural next target.

\bibliographystyle{ACM-Reference-Format}
\bibliography{refs}

\clearpage

\section*{Appendix: Full Rule Listings}

\begin{figure}[h]
\begin{lstlisting}[language=datalog,moredelim={[is][\bfseries]{|}{|}}]
// Base heap edges through array slots, then a linear TC.
HeapReachable0(b, h) :- ArrayIndexPointsTo(b, h).
HeapReachable(b, h) :- HeapReachable0(b, h).
HeapReachable(b, h) :- |HeapReachable(b, m)|, HeapReachable0(m, h).
// Global escape: anything stored in a static field, then
// closed over heap reachability.
GlobalEscape(h) :- StaticFieldPointsTo(h, _).
GlobalEscape(h) :- |GlobalEscape(b)|, HeapReachable(b, h).
// Per-method escape (five cases): (1) globally escaped, 
// (2) returned, (3) passed as actual, (4) used as receiver, 
// or (5) reachable from another escaping alloc.
MethodEscape(h, m) :- AllocatedIn(h, m), GlobalEscape(h).
MethodEscape(h, m) :-
  AllocatedIn(h, m), ReturnVar(v, m), VarPointsTo(h, v).
MethodEscape(h, m) :-
  AllocatedIn(h, m), CallActualIn(m, v), VarPointsTo(h, v).
MethodEscape(h, m) :-
  AllocatedIn(h, m), CallRecvIn(m, v), VarPointsTo(h, v).
MethodEscape(h, m) :-
  AllocatedIn(h, m), AllocatedIn(o, m),
  |MethodEscape(o, m)|, HeapReachable(o, h).
// Captured: allocated in a reachable method, never escapes.
CapturedAllocation(h, m) :-
  AllocatedIn(h, m), Reachable(m), !MethodEscape(h, m).
\end{lstlisting}
\caption{Escape analysis (\texttt{escape}). \textit{HeapReachable}, \textit{GlobalEscape}, and \textit{MethodEscape} are each linear-recursive; the recursive literal of each rule is shown in bold. The listing shows all 11 rules and all five \textsc{IDB}\xspace predicates counted in Table~\ref{tab:bench-summary}; the points-to summaries (\textit{VarPointsTo}, \textit{StaticFieldPointsTo}, and \textit{ArrayIndexPointsTo}) are precomputed inputs and thus \textsc{EDB}\xspace relations here.}
\label{fig:escape}
\end{figure}

\begin{figure}[h]
\begin{lstlisting}[language=datalog,moredelim={[is][\bfseries]{|}{|}}]
// Linear TC of child_path.
ancestor_path(x, y) :- child_path(x, y).
ancestor_path(x, y) :- 
  |ancestor_path(z, y)|, child_path(z, x).
// Move/assignment information propagates down ancestor_path
path_moved_at(x, y) :- path_moved_at_base(x, y).
path_moved_at(x, y) :- 
  |path_moved_at(z, y)|, ancestor_path(z, x).
path_assigned_at(x, y) :- path_assigned_at_base(x, y).
path_assigned_at(x, y) :- 
  |path_assigned_at(z, y)|, ancestor_path(z, x).
// Flow moves along the CFG, killed by an intervening 
// assignment (stratified negation).
path_maybe_uninitialized_on_exit(p, p2) :- 
  path_moved_at(p, p2).
path_maybe_uninitialized_on_exit(p, p2) :-
  |path_maybe_uninitialized_on_exit(p, p1)|,
  cfg_edge(p1, p2), !path_assigned_at(p, p2).
// A use of a maybe-uninitialized path is an error.
move_error(p, t) :-
  path_maybe_uninitialized_on_exit(p, s), cfg_edge(s, t).
\end{lstlisting}
\caption{Polonius borrow checker (\texttt{borrow}). Each recursive rule contains exactly one positive recursive literal, shown in bold; negation appears only at \textsc{SCC}\xspace boundaries.}
\label{fig:polonius}
\end{figure}

\begin{figure}[h]
\begin{lstlisting}[language=datalog,moredelim={[is][\bfseries]{|}{|}}]
VarPointsTo(h, v) :- AssignHeapAlloc(h, v, m), Reachable(m).
VarPointsTo(h, to) :- Assign(f, to), |VarPointsTo(h, f)|.
VarPointsTo(h, to) :- Reachable(m),
  AssignLocal(from, to, m), |VarPointsTo(h, from)|.
VarPointsTo(h, to) :- Reachable(m), AssignCast(t, f, to, m),
  SupertypeOf(t, a), HeapAllocType(h, a), |VarPointsTo(h, f)|.
VarPointsTo(h, o) :- Reachable(m), LoadArrayIndex(b, o, m),
  |VarPointsTo(bh, b)|, ArrayIndexPointsTo(bh, h),
  VarType(o, t), HeapAllocType(bh, bt),
  ComponentType(bt, ct), SupertypeOf(t, ct).
VarPointsTo(h, o) :- Reachable(m), 
  LoadInstanceField(b, s, o, m),
  |VarPointsTo(bh, b)|, InstanceFieldPointsTo(h, s, bh).
VarPointsTo(h, to) :- Reachable(m),
  LoadStaticField(fld, to, m), |StaticFieldPointsTo(h, fld)|.
VarPointsTo(h, this) :- Reachable(m), InstrMethod(inv, m),
  VirtualCallBase(inv, base), |VarPointsTo(h, base)|,
  HeapAllocType(h, ht), VirtualCallName(inv, n), VirtualCallDesc(inv, d),
  MethodLookup(n, d, ht, tm), ThisVar(tm, this).
VarPointsTo(h, this) :- Reachable(m), InstrMethod(inv, m),
  SpecialCallBase(inv, base), |VarPointsTo(h, base)|,
  CallTarget(inv, tm), ThisVar(tm, this).
StaticFieldPointsTo(h, fld) :- Reachable(m),
  StoreStaticField(from, fld, m), |VarPointsTo(h, from)|.
\end{lstlisting}
\caption{The complete ten-rule \texttt{varpointsto} program, of which Figure~\ref{fig:pts-rules} shows four rules. The recursive literal of each rule is shown in bold: \textit{VarPointsTo} and \textit{StaticFieldPointsTo} form a single mutually recursive \textsc{SCC}\xspace, and every rule contains at most one positive literal from this \textsc{SCC}\xspace, so the program is linear. Variable and \textsc{EDB}\xspace predicate names are abbreviated relative to the Soufflé source.}
\label{fig:varpointsto-full}
\end{figure}

\clearpage

\end{document}